\documentclass[review]{elsarticle}

\usepackage{lineno,hyperref}
\modulolinenumbers[5]

\journal{Annals of Physics}

\usepackage{amsmath,amssymb,physics,color}

\newcommand{\thop}{t_\textrm{h}}
\newcommand{\Crnd}{C^\textrm{rnd}}

\begin{document}

\begin{frontmatter}

\title{Delocalization of non-Hermitian Quantum Walk on Random Media in One Dimension}

\author{Naomichi Hatano}
\address{Institute of Industrial Science, The University of Tokyo, 5-1-5 Kashiwanoha, Kashiwa, Chiba 277-8574, Japan}
\ead{hatano@iis.u-tokyo.ac.jp}

\author{Hideaki Obuse}
\address{Department of Applied Physics, Hokkaido University, Sapporo 060-8628, Japan}
\address{Institute of Industrial Science, The University of Tokyo, 5-1-5 Kashiwanoha, Kashiwa, Chiba 277-8574, Japan}
\ead{hideaki.obuse@eng.hokudai.ac.jp}





\begin{abstract}
Delocalization transition is numerically found in a non-Hermitian extension of a discrete-time quantum walk on a one-dimensional random medium.
At the transition, an eigenvector gets delocalized and at the same time the corresponding energy eigenvalue (the imaginary unit times the phase of the eigenvalue of the time-evolution operator) becomes complex.
This is in accordance with a non-Hermitian extension of the random Anderson model in one dimension, called, the Hatano-Nelson model.
We thereby numerically find that all eigenstates of the Hermitian quantum walk share a common localization length.
\end{abstract}


\end{frontmatter}

\linenumbers

\section{Introduction}
\label{sec1}

Non-Hermiticity in quantum mechanics is attracting much attention in various fields.
The non-Hermiticity was presumably used for the first time in history of quantum mechanics in the field of nuclear physics.
We can perhaps go back to Gamow~\cite{Gamow28}, who tried to explain resonant scattering in terms of resonant states with complex eigenvalues, which indeed emerge in open quantum systems.
Several papers on resonant states follow intermittently~\cite{Siegert39,Peierls59,leCouteur60}.
Perhaps Feshbach played the most decisive role in the field of nuclear physics~\cite{Feshbach58,Feshbach62,Feshbach58review}; 
he introduced the optical model, which has a complex potential, and then justified it in terms of a theory in which he eliminated the environmental degrees of freedom to obtain an effective complex potential.
This is indeed a theory of open quantum systems in the present-day terminology.
This approach continues further onto the field of quantum statistical physics; see Ref.~\cite{Breuer-Petruccione} for a good textbook.

In late 1990s, there  independently appeared three pieces of work that shifted the paradigm of non-Hermitian quantum mechanics; namely a non-Hermitian random Anderson model (often called the Hatano-Nelson model)~\cite{HN96} in 1996, the non-Hermiticity in stochastic processes~\cite{Chalker97} in 1997, and the $PT$-symmetric theory~\cite{Bender98} in 1998.
Although the original motivation of introducing the non-Hermiticity varied, they all analyzed novel types of non-Hermiticity and stimulated many researches that was motivated to study the non-Hermiticity itself rather than stumbling on an effective non-Hermitian model.
These days, there appear many experiments that try to materialize theoretically proposed incidents of the non-Hermiticity\cite{Guo09,Rueter10,Peng14,Feng14,Hodaei14}.


The main purpose of the present paper is to introduce a non-Hermitian discrete-time quantum walk on a one-dimensional random medium and to show numerically that it exhibits a localization-delocalization transition in the same manner as the Hatano-Nelson model.
In the Hatano-Nelson model, two things happen at the same time at the transition point:
first, an eigenvector that is localized because of the Anderson localization in one dimension gets delocalized;
second, the corresponding energy eigenvalue becomes complex.
While the delocalization transition point depends on the energy in the original Hatano-Nelson model,
we here find for our non-Hermitian quantum walk that all eigenstates simultaneously undergoes the non-Hermitian delocalization transition when we turn up a non-Hermitian parameter, which implies that all eigenstates of the Hermitian quantum walk have a common localization length. 
We argue that this observation results from the absence of symmetry of the Hermitian quantum walk and the periodicity of the energy eigenvalue without band gaps.
(In the latter, the energy eigenvalue means the imaginary unit times the phase of the eigenvalue of the time-evolution operator.)

We review in Sec.~\ref{sec2} this transition in the Hatano-Nelson model.
Section~\ref{sec3} presents our new results for the non-Hermitian quantum walk.
Section~\ref{sec4} is devoted to a summary.

\section{An overview of the Hatano-Nelson model}
\label{sec2}

For later use in Sec.~\ref{sec3} for the definition of non-Hermitian quantum walk on random media, we here present a brief overview of a non-Hermitian random Anderson model.
In 1996, one of the present authors (N.H.) together with a collaborator introduced a non-Hermitian extension of the random Anderson model, which are now often referred to as the Hatano-Nelson model~\cite{HN96,HN97}:
\begin{align}\label{eq10}
H\qty(\vec{g}):=\frac{\qty(\vec{p}+i\vec{g})^2}{2m}+V\qty(\vec{x}),
\end{align}
where $\vec{p}$ is the momentum operator, $\vec{g}$ is a constant real vector, which we refer to as the imaginary vector potential, and $V\qty(\vec{x})$ is a real random potential.
Note that the Hamiltonian is non-Hermitian $H^\dag\neq H$ when $\vec{g}\neq\vec{0}$.
This was one of the first studies on non-Hermitian systems that emerged in the late 1990s~\cite{Chalker97,Bender98}.

Throughout the present paper, we will refer to its one-dimensional lattice version~\cite{HN96,HN97}:
\begin{align}\label{eq20}
H(g):=-\thop\sum_{x=-\infty}^\infty
\qty(e^g \dyad{x+1}{x}+e^{-g}\dyad{x}{x+1})
+\sum_{x=-\infty}^\infty V_x\dyad{x},
\end{align}
where $\thop$ is the tight-binding hopping element, which for brevity we set to unity hereafter, $g$ is a lattice version of the imaginary vector potential with the unit $\hbar=1$, and $V_x$ is a site-random real potential.
If $i\vec{g}$ in Eq.~\eqref{eq10} were a real gauge field $e\vec{A}$, it would be translated to phase factors $e^{\pm ieA}$ in the lattice Hamiltonian~\eqref{eq20} according to the Peierls substitution.
In the non-Hermitian Hamiltonian~\eqref{eq20}, we replace the phase factors $e^{\pm ieA}$ by the amplitude modulations $e^{\mp g}$.
We remark that the non-Hermitial Hamiltonian possesses time-reversal symmetry defined by $H(g)=H^*(g)$~\cite{Kawabata19}.

Although the Hatano-Nelson model was originally introduced after an inverse path-integral mapping of a statistical-physical model of type-II superconductors with columnar defects and a magnetic flux, the resulting quantum model had an interesting implication for the Hermitian random Anderson model.
%
Let us explain it for the lattice Hamiltonian~\eqref{eq20} in one dimension.
If $g$ in Eq.~\eqref{eq20} were $ieA$, we would be able to gauge it out completely and eliminate the vector potential by means of the gauge transformation $\ket{x}\to e^{ieAx}\ket{x}$, which would modify only the phase of the eigenvectors.
We could similarly use a transformation, which we refer to as the imaginary gauge transformation, of the form
\begin{align}\label{eq30}
W\ket{x} &= e^{gx}\ket{x},
\\\label{eq40}
\bra{x}W^{-1} &= e^{-gx}\bra{x},
\end{align}
and thereby reduce the non-Hermitian Hamiltonian $H(g)$ to the Hermitian limit $H(0)$.
It might be easier to understand it in the matrix representation.
While the Hamiltonian~\eqref{eq20} is given in the evidently non-Hermitian form
\begin{align}
H(g)=\mqty( 
\ddots & \ddots & & & & & \\
\ddots & V_{x-2} & -e^{-g} & & & &\\
& -e^g & V_{x-1} & -e^{-g} & & & \\
& & -e^g & V_x & -e^{-g} & & \\
& & & -e^g & V_{x+1} & -e^{-g} & \\
& & & & -e^g & V_{x+2} & \ddots \\
& & & & & \ddots & \ddots 
),
\end{align}
the imaginary gauge transformation~\eqref{eq30}--\eqref{eq40} is not a unitary transformation but a similarity transformation
\begin{align}
W=\mqty(
\ddots & & & & & & \\
& e^{g(x-2)} & & & & & \\
& & e^{g(x-1)} & & & & \\
& & & e^{gx} & & & \\
& & & & e^{g(x+1)} & & \\
& & & & & e^{g(x+2)} & \\
& & & & & & \ddots 
).s
\end{align}
They would produce
\begin{align}
\qty(W^{-1}H(g)W)_{x,x}&=\qty(W^{-1})_{x,x}\qty(H(g))_{x,x}\qty(W)_{x,x}=V_x,
\\
\qty(W^{-1}H(g)W)_{x+1,x}&=\qty(W^{-1})_{x+1,x+1}\qty(H(g))_{x+1,x}\qty(W)_{x,x}
\nonumber\\
&=-e^{-g(x+1)} e^g e^{gx}=-1,
\\
\qty(W^{-1}H(g)W)_{x,x+1}&=\qty(W^{-1})_{x,x}\qty(H(g))_{x,x+1}\qty(W)_{x+1,x+1}
\nonumber\\
&=-e^{-gx} e^{-g} e^{g(x+1)}=-1,
\end{align}
and hence we would find 
\begin{align}\label{eq95}
W^{-1}H(g)W=H(0).
\end{align}
Since we can transform the non-Hermitian Hamiltonian $H(g)$ to the Hermitian one $H(0)$ under a similarity transformation, we would conclude that the Hermitian spectrum with all real eigenvalues is common to the non-Hermitian Hamiltonian for any $g$.

In fact, this is not true if we restrict ourselves to the Hilbert space.
Each real eigenvalue survives only for a specific region of small values of $g$.
In order to explain it, let us express an eigenvector of the Hamiltonian $H(0)$ with 
\begin{align}
H(0)\ket{\psi_n(0)}=E_n\ket{\psi_n(0)}
\end{align} 
in the form
\begin{align}\label{eq100}
\ket{\psi_n(0)}=\sum_{x=-\infty}^\infty c^{(n)}_x(0)\ket{x}.
\end{align}
Since we have a site-random real potential $V_x$ in one dimension, the argument of the Anderson localization dictates that any eigenvector is localized in the form
\begin{align}\label{eq110}
\abs{c^{(n)}_x(0)}\sim e^{-\kappa_n \abs{x-x^{(n)}_\textrm{c}}},
\end{align}
where $\kappa_n$ and $x^{(n)}_\textrm{c}$ are the inverse localization length and the localization center, respectively, of the $n$th eigenvector.
Following the argument of the imaginary gauge transformation~\eqref{eq30}--\eqref{eq40}, we would find the corresponding $n$th right-eigenvector of $H(g)$ in the form 
\begin{align}\label{eq115}
\ket{\psi^\textrm{R}_n(g)}:=W\ket{\psi_n(0)}
\end{align}
because we have 
$H(g)W\ket{\psi_n(0)}=E_nW\ket{\psi_n(0)}$
from Eq.~\eqref{eq95}.
Note here that since $H(g)$ for $g\neq0$ is non-Hermitian, the left-eigenvector defined as the solution of
\begin{align}
\bra{\psi^\textrm{L}_n(g)}H(g)=E_n\bra{\psi^\textrm{L}_n(g)}
\end{align}
is generally \textit{not} the Hermitian conjugate of the right-eigenvector:
$\bra{\psi^\textrm{L}_n(g)}\neq \ket{\psi^\textrm{R}_n(g)}^\dag$.
Similarly to the right-eigenvector~\eqref{eq115}, we would obtain the $n$th left-eigenvector in the form
\begin{align}\label{eq118}
\bra{\psi^\textrm{L}_n(g)}=\bra{\psi_n(0)}W^{-1}
\end{align}
because we have $\bra{\psi_n(0)}W^{-1}H(0)=E_n\bra{\psi_n(0)}W^{-1}$ from Eq.~\eqref{eq95}.

We would thereby conclude for the non-Hermitian Hamiltonian $H(g)$ that the eigenvalue remains the same $E_n$ for \textit{any} value of $g$ and the right-eigenvector is modified in the form
\begin{align}\label{eq130}
\ket{\psi_n(g)}=\sum_{x=-\infty}^\infty c^{(n)}_x(g)\ket{x}
\end{align}
with 
\begin{align}\label{eq140}
\abs{c^{(n)}_x(g)}\sim e^{-\kappa_n \abs{x-x^{(n)}_\textrm{c}}+gx},
\end{align}
which is \textit{not} normalizable if $\abs{g}>\kappa_n$, however, and hence does not belong to the Hilbert space.
The correct conclusion is that we can apply the imaginary gauge transformation~\eqref{eq30}--\eqref{eq40} to the $n$th eigenvector~\eqref{eq100}--\eqref{eq110} as long as $\abs{g}<\kappa_n$, and hence the eigenvalue $E_n$ is fixed only in this regime.
Once $\abs{g}$ exceeds $\kappa_n$, the argument of the imaginary gauge transformation does not tell us anything.

So far, we have considered the infinite system.
In reality, we would numerically analyze the model of finite size.
Indeed, we will see below that a numerical analysis for systems under a periodic boundary condition gives us a hint for what happens in the region $\abs{g}>\kappa_n$.
Nonetheless, we should note that the conclusion from the numerical analysis of finite systems is very much different for periodic systems and open systems.

\begin{figure}
\centering
\includegraphics[width=0.5\textwidth]{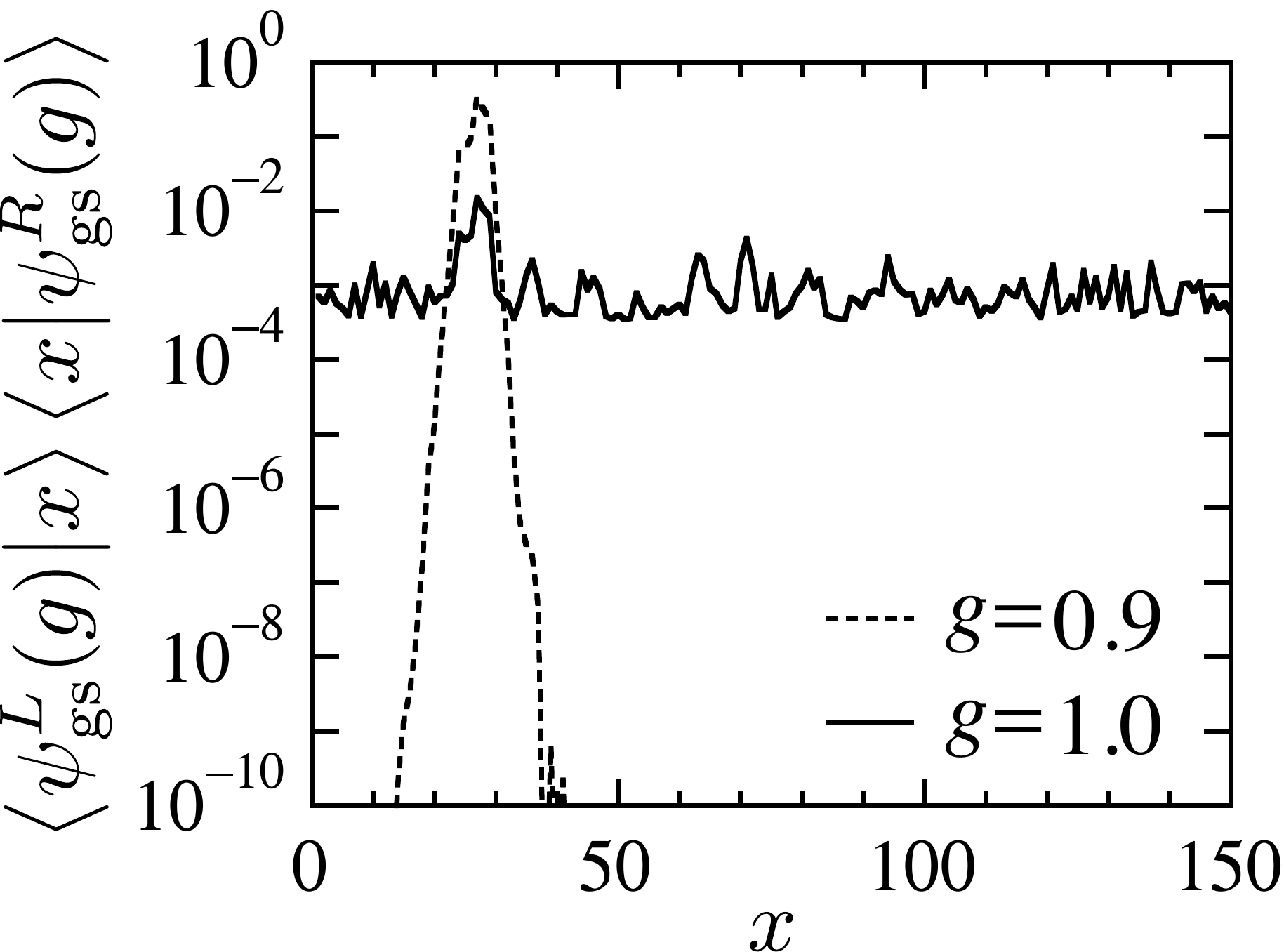}
\caption{The product of the right- and left-eigenvectors (the squared norm of the eigenvector) for the lowest eigenvalue of the lattice Hamiltonian~\eqref{eq20} of size $L=1000$;
the part of $0\leq x \leq150$ is shown.
The site-random potential is set for each site from the range $[-1,1]$.
The broken line indicates the product for $g=0.9$, while the solid line for $g=1.0$.}
\label{fig1}
\end{figure}

For periodic systems of size $L$, the argument for the imaginary gauge transformation is valid again in the regime of $\abs{g}<\kappa_n$ only, at least for large systems.
The gauge-transformed $n$th eigenvector~\eqref{eq130}--\eqref{eq140} would be almost consistent with the periodicity as long as $\abs{g}<\kappa_n$;
the contradiction at the points $x=x_\textrm{c}+L/2$ and $x=x_\textrm{c}-L/2$ between $\exp[-(\kappa_n-\abs{g})L/2]$ and $\exp[-(\kappa_n+\abs{g})L/2]$ should be exponentially small as long as $(\kappa_n-\abs{g})L\gg 1$.
The argument of the imaginary vector potential is gradually violated as $\abs{g}$ approaches $(\kappa-1/L)$ from below and absolutely invalidated when $\abs{g}>\kappa_n$.
In this sense, large periodic systems mimic the infinite system in the Hilbert space.
We exemplify this in Fig.~\ref{fig1}, where we plotted the product of the right- and left-eigenvector of the ground state, which is defined to have the lowest real part of the eigenvalue.
The argument of the imaginary gauge transformation would result in the conclusion that the product of Eqs.~\eqref{eq115} and~\eqref{eq118} is conserved in the following sense:
\begin{align}\label{eq190}
\braket{\psi^\textrm{L}_n(g)}{\psi^\textrm{R}_n(g)}=\braket{\psi_n(0)}{\psi_n(0)}
\end{align}
for any value of $g$.
In reality, the eigenvector in Fig.~\ref{fig1} drastically changes at a point in the region $g=[0.9,1.0]$, from which we can presume that the inverse localization length $\kappa_n$ of this particular localized eigenvector is somewhere between $0.9$ and $1.0$.
When $g=0.9$, we have $|g|<\kappa_n$, and hence we indeed find Eq.~\eqref{eq190}, but when $g=1.0$, the eigenvector seems delocalized in contradiction with the common knowledge that all eigenvectors are localized in one-dimensional random media.

\begin{figure}
\begin{minipage}[t]{0.48\textwidth}
\vspace{0pt}
\includegraphics[width=\textwidth]{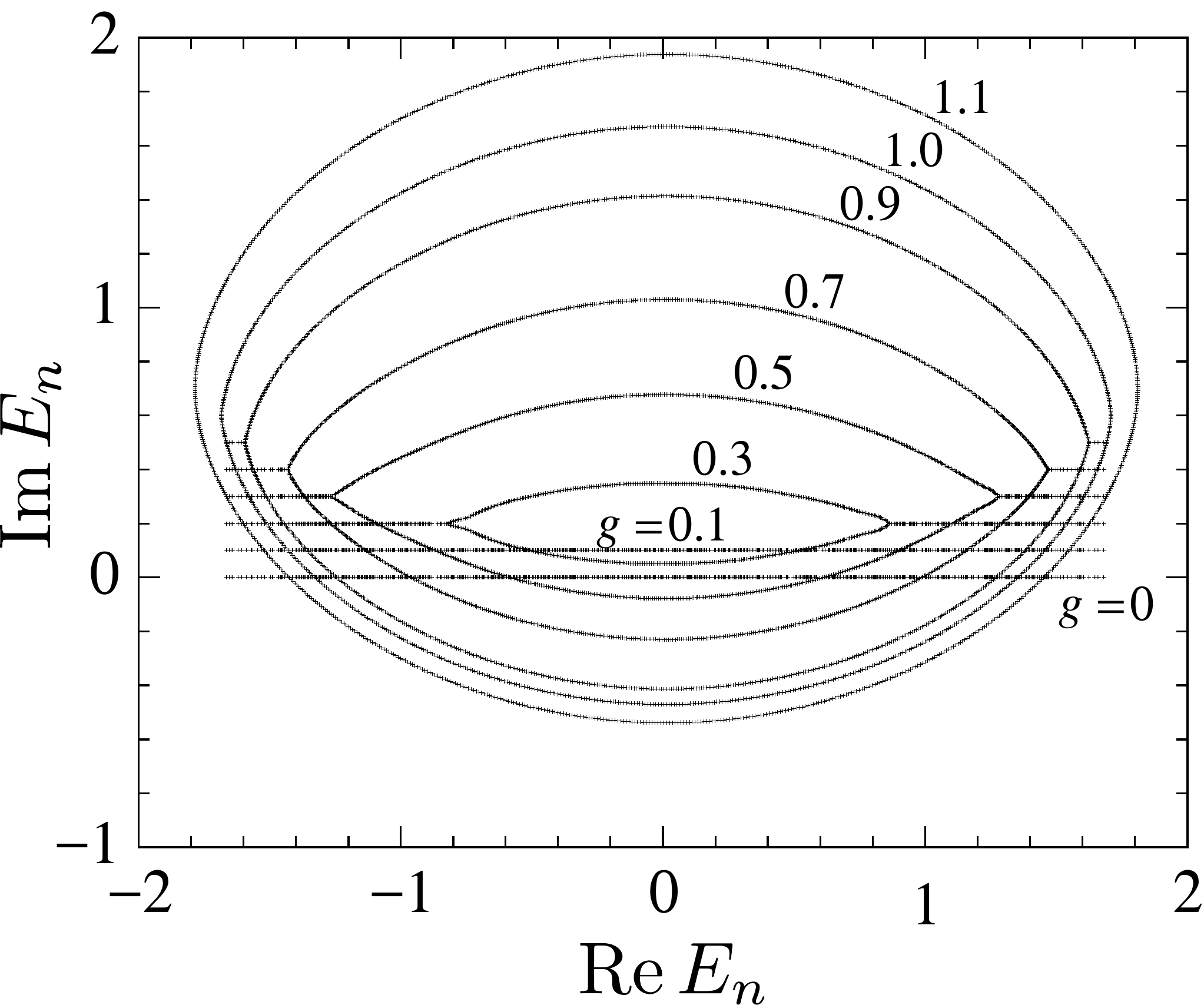}
\centering \footnotesize{(a)}
\end{minipage}
\hfill
\begin{minipage}[t]{0.48\textwidth}
\vspace{5pt}
\includegraphics[width=\textwidth]{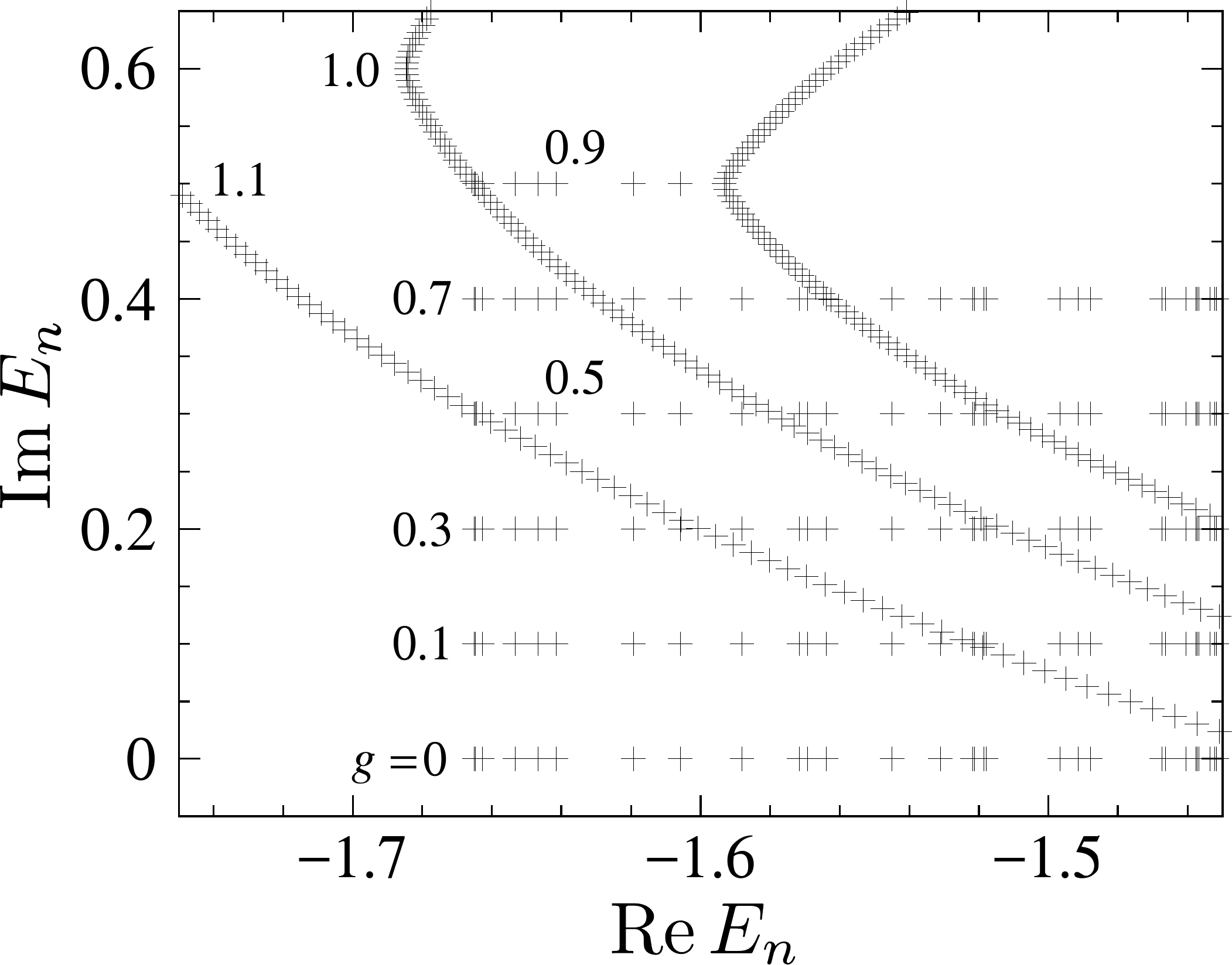}
\centering \footnotesize{(a)}
\end{minipage}
\caption{The eigenvalue distribution of the lattice Hamiltonian $H(g)$ of size $L=1000$ under the periodic boundary condition. The site-random potential is set for each site from the range $[-1,1]$, while the non-Hermitian parameter $g$ is varied from $0$ to $1.1$.
Each set of the eigenvalues for the respective value of $g$ is symmetric with respect to the real axis,
but it is shifted up with a respective offset so that it may not overlap with the other sets. (b) is an enhanced view of a part of (a).
Taken from Refs.~\cite{HN96,HN97}.}
\label{fig2}
\end{figure}

\begin{figure}
\centering
\includegraphics[width=0.6\textwidth]{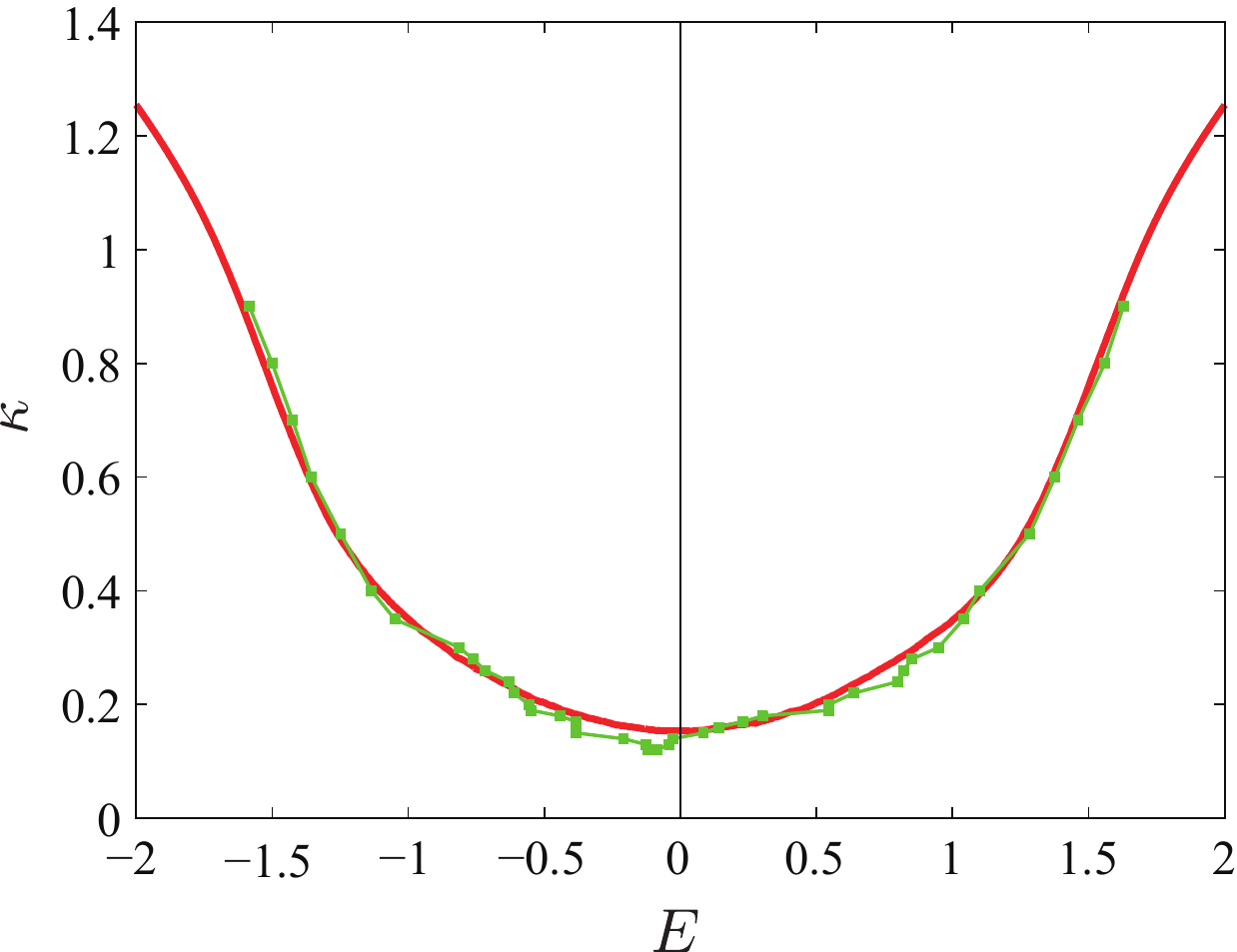}
\caption{Energy dependence of the inverse localization length $\kappa$ of the Hermitian random Anderson model $H(0)$.
The site-random potential is set for each site from the range $[-1,1]$.
Green dots indicate the estimates from the variation of the energy spectrum shown in Fig.~\ref{fig2}(a), while the solid red curve indicate an estimate by the Chebyshev-polynomial expansion found in Ref.~\cite{HF16}.
Taken from Ref.~\cite{HF16}.}
\label{fig3}
\end{figure}

The eigenvalue distribution changes as we vary the value of $g$ as in Fig.~\ref{fig2}(a).
We can understand this variation as follows~\cite{HN96,HN97}.
For the Hermitian random Anderson model $H(0)$, the inverse localization length depends on the energy as exemplified in Fig.~\ref{fig3}.
As we vary $g$ from $0$ up to $0.1$, all eigenvectors satisfy the inequality $\abs{g}<\kappa_n$;
hence the imaginary gauge transformation is applicable to all states, and all eigenvalues remain fixed.
This is what we observe in Fig.~\ref{fig2}(a) and more closely in Fig.~\ref{fig2}(b).
As we increase $g$ further, the inequality $\abs{g}<\kappa_n$ is violated for states around the center of the energy spectrum.
Accordingly, we indeed observe in Fig.~\ref{fig2}(a) that the eigenvalues around the center are not fixed anymore; 
they become pairs of complex eigenvalues.
On the other hand, the inequality $\abs{g}<\kappa_n$ is \textit{not} violated even for the same value of $g$ for states closer to the spectrum edges.
Therefore, their eigenvalues are fixed to the original real ones, as we can observe in Fig.~\ref{fig2}(b).
Since the energy range where the inequality $\abs{g}<\kappa_n$ is violated becomes wider and wider as we further turn up $g$, the range of the complex eigenvalues also expand accordingly.
By inverting the logic, we realize that we should find the equality $\abs{g}=\kappa_n$ at the edges of the energy range of the complex eigenvalues;
this is indeed how Ref.~\cite{HN97} estimated the energy dependence of the inverse localization length as indicated by dots in Fig.~\ref{fig3}.

To summarize the argument for periodic finite systems, two transitions occur at the same time when $g=\kappa_n$:
first, the delocalization transition occurs for the $n$th eigenvector;
second, the corresponding eigenvalue $E_n$ becomes complex.

For open finite systems, on the other hand, the argument of the imaginary gauge transformation is always valid because the eigenvector is normalizable for any value of $g$.
For a positive value of $g$, the right-eigenvectors is exponentially large on the right edge, whereas for a negative value of $g$, the left-eigenvector is on the left edge.
Nevertheless, they are still normalizable because of the finite size of the system.
Therefore, all eigenvalues remain the same real values.
This is recently called the non-Hermitian skin effect in the literature~\cite{YaoWang18},
but we will not go into its details since it is out of the scope of the present paper.

\section{Delocalization Transition of Non-Hermitian Quantum Walk on a Random Chain}
\label{sec3}

The purpose of the present paper is to convert the argument around the delocalization transition in Sec.~\ref{sec2} for the non-Hermitian random Anderson model into the one for a non-Hermitian extension of the discrete-time quantum walk on random media.
We first review the standard discrete-time quantum walk in one dimension, then define a non-Hermitian quantum walk, and finally demonstrate its delocalization transition numerically.
We thereby find that all eigenstates of the Hermitian quantum walk has a common localization length.

The standard quantum walk in one dimension is defined on a chain in which each site accommodates two states as an inner degree of freedom, namely the left mover and the right mover, which we denote by
$\ket{xL}$ and $\ket{xR}$ for a site $x$, respectively.
Let us first introduce the shift operator $S$, which work on the states as
\begin{align}
S\ket{xL}&=\ket{(x-1)L},
\\
S\ket{xR}&=\ket{(x+1)R}
\end{align}
for any site $x$.
In the matrix representation, we can express it in the form
\begin{align}
S=
\left(
\begin{array}{cccccccccccc}
\multicolumn{1}{c|}{\ddots} &&&&&&&&&&& \\
\cline{1-3}
&\multicolumn{1}{|c}{0} &\multicolumn{1}{c|}{0} &1& &&&&&&& \\
1&\multicolumn{1}{|c}{0} &\multicolumn{1}{c|}{0} && &&&&&&& \\
\cline{2-5}
&&& \multicolumn{1}{|c}{0} & \multicolumn{1}{c|}{0} & 1& & &&&&\\
&&1& \multicolumn{1}{|c}{0} & \multicolumn{1}{c|}{0} & & & &&&&\\
\cline{4-7}
&&& & & \multicolumn{1}{|c}{0} & \multicolumn{1}{c|}{0} &1 &&&&\\
&&& & 1& \multicolumn{1}{|c}{0} & \multicolumn{1}{c|}{0} & &&&&\\
\cline{6-9}
&&&&&&& \multicolumn{1}{|c}{0}& \multicolumn{1}{c|}{0} &1&&\\
&&&&&&1& \multicolumn{1}{|c}{0}& \multicolumn{1}{c|}{0} &&&\\
\cline{8-11}
&&&&&&&&& \multicolumn{1}{|c}{0}& \multicolumn{1}{c|}{0} &1\\
&&&&&&&&1& \multicolumn{1}{|c}{0}& \multicolumn{1}{c|}{0} &\\
\cline{10-12}
&&&&&&&&&&& \multicolumn{1}{|c}{\ddots}
\end{array}
\right),
\end{align}
using the basis set 
\begin{align}
\cdots,
\ket{(x-2)L},
\ket{(x-2)R} ,
&\ket{(x-1)L} ,
\ket{(x-1)R} ,
\ket{xL} ,
\ket{xR} ,
\nonumber\\
&\ket{(x+1)L} ,
\ket{(x+1)R} ,
\ket{(x+2)L} ,
\ket{(x+2)R},
\cdots.
\end{align}
Note that this is a unitary operator.

If we had only the shift operator, the right mover would keep moving to the right ballistically, the left mover would keep moving to the left ballistically, and nothing else would happen.
We next introduce the coin operator $C$ to shuffle the ballistic movements of the left and right movers at each site:
\begin{align}\label{eq240}
C
\mqty( \ket{xL} \\ \ket{xR} )
=\mqty( \alpha & \gamma \\
\beta & \delta)
\mqty( \ket{xL} \\ \ket{xR} )
\end{align}
for any site $x$,
where the two-by-two matrix on the right-hand side is a unitary matrix, and hence $\abs{\alpha}^2+\abs{\beta}^2=\abs{\gamma}^2+\abs{\delta}^2=1$ with $\alpha^\ast\gamma+\beta^\ast\delta=0$.
The time evolution of a quantum walk consists of operating $S$ and $C$ on the initial state alternatively, as is expressed by a unitary operator $U=SC$.

\begin{figure}
\centering
\includegraphics[width=0.75\textwidth]{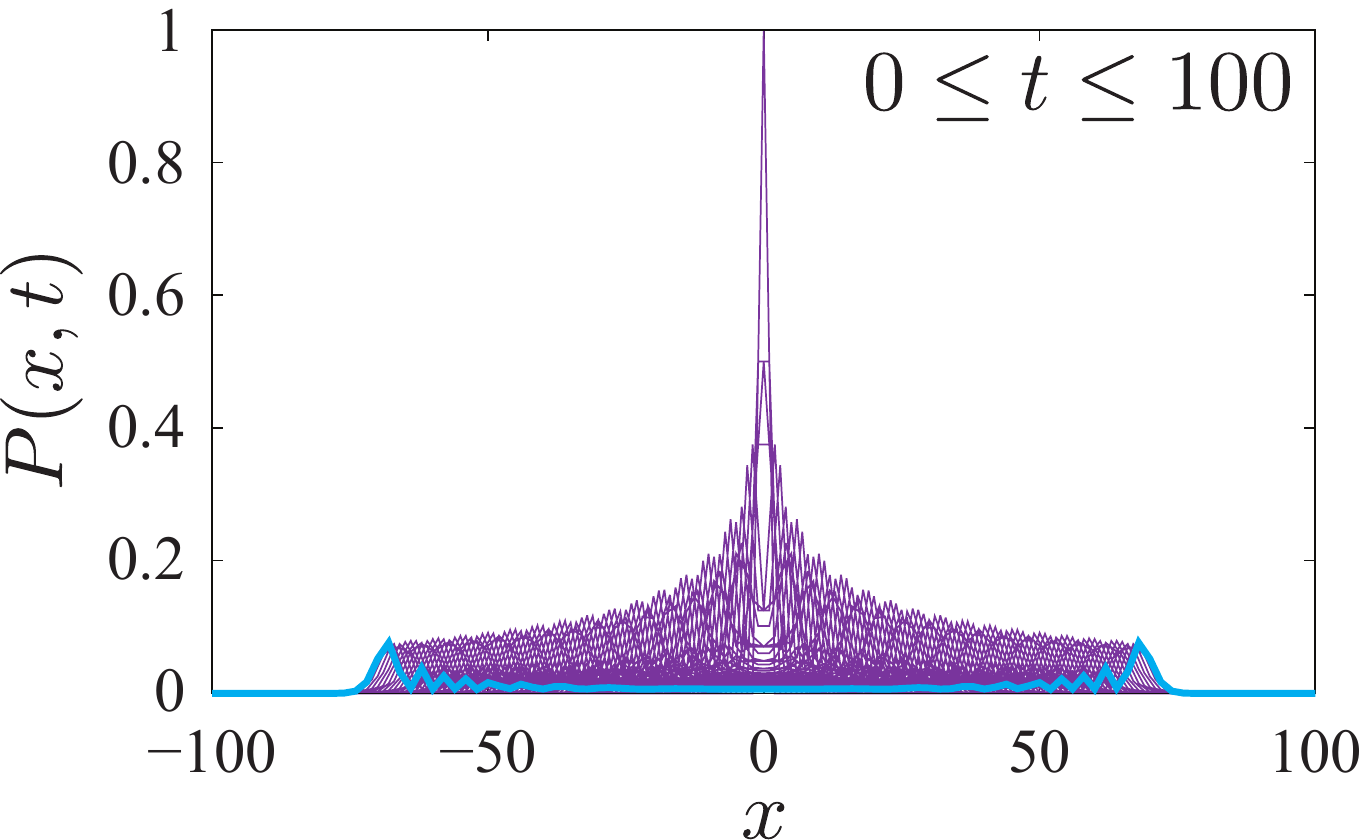}
\\
\footnotesize{(a)}
\\[12pt]
\includegraphics[width=0.75\textwidth]{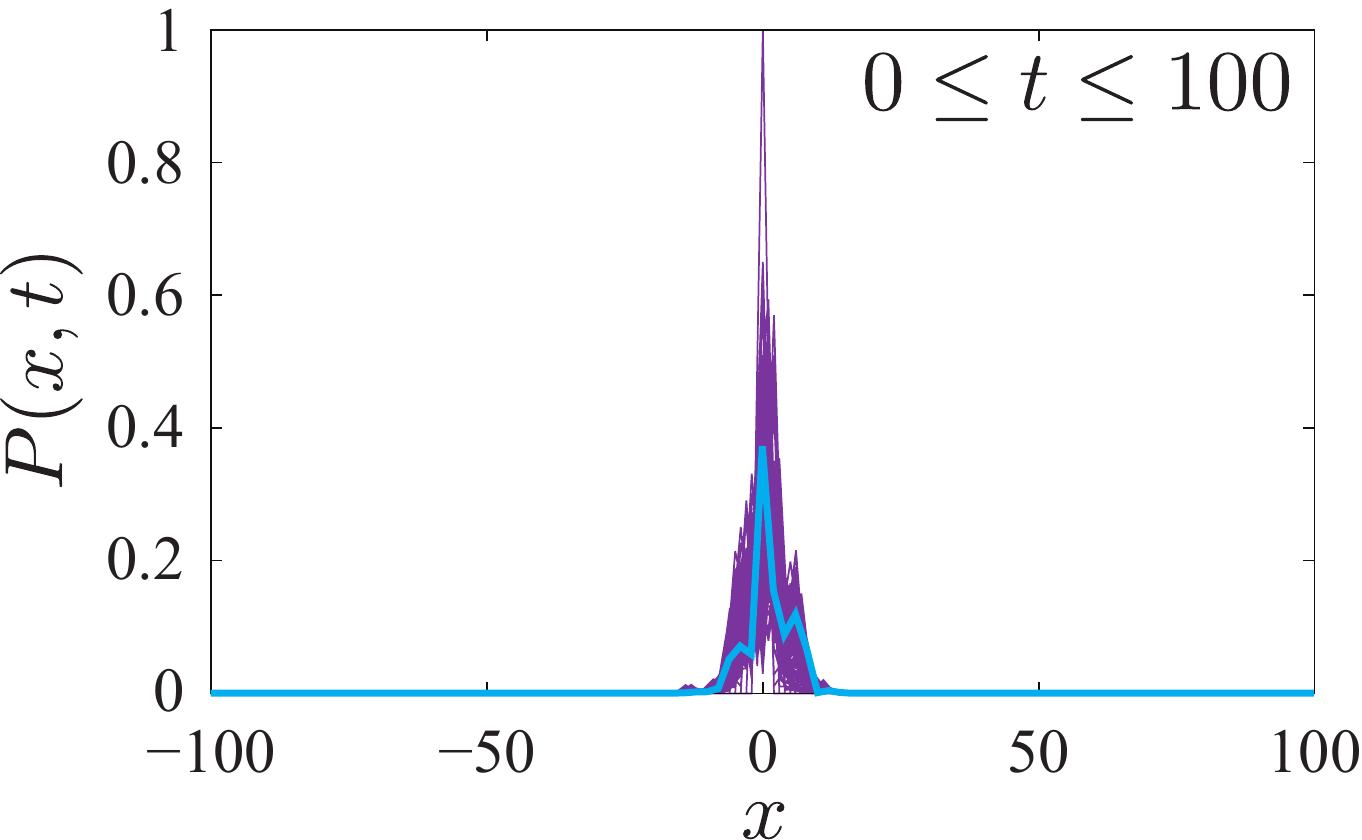}
\\
\footnotesize{(b)}
\caption{Time evolution of the discrete-time quantum walk with (a) the uniform coin operator and (b) the random coin operator. In each panel, the thin purple lines indicate the profiles of the probability amplitude from $t=0$ to $t=99$, while the thick blue line indicate the last one at $t=100$.}
\label{fig4}
\end{figure}

We exemplify the time evolution in Fig.~\ref{fig4}(a).
We here plotted the probability amplitude 
\begin{align}\label{eq250}
P(x,t):=\abs{\braket{xL}{\psi(t)}}^2+\abs{\braket{xR}{\psi(t)}}^2,
\end{align}
where
\begin{align}
\ket{\psi(t)}=(SC)^t\ket{\psi(0)}
\end{align}
with the coin operator set to
\begin{align}
C=\frac{1}{\sqrt{2}}\mqty(
1 & 1 \\
-1 & 1 
)
\label{eq:Csimple}
\end{align}
and the initial state set to
\begin{align}
\ket{\psi(0)}=\frac{1}{\sqrt{2}}\qty(\ket{0L}+i\ket{0R}).
\end{align}
We can observe a ballistic propagation of the wave fronts, which is one of the critical features of the discrete-time quantum walk.

We can introduce a site-random potential by making the two-by-two matrix in Eq.~\eqref{eq240} for each site a random unitary matrix:
\begin{align}\label{eq290}
\Crnd
\mqty( \ket{xL} \\ \ket{xR} )
=\Crnd_x\mqty( \ket{xL} \\ \ket{xR}),
\end{align}
where $\Crnd_x$ is a random unitary matrix
\begin{align}
\Crnd_x :=
e^{i\phi}
\begin{pmatrix}
e^{i\alpha} \cos \vartheta & -e^{i \beta} \sin \vartheta \\
e^{-i\beta} \sin \vartheta & e^{-i \alpha} \cos \vartheta
\end{pmatrix},
\label{eq:U2}
\end{align}
where the values of $\alpha$, $\beta$, $\phi$, and  $\vartheta$ are chosen independently for each site $x$ from an arbitrary ensemble.
The randomness generally localizes the quantum walker, as is exemplified in Fig.~\ref{fig4}(b).
We here plotted the probability amplitude~\eqref{eq250} with the coin operator replaced by $\Crnd$ as in
\begin{align}
\ket{\psi(t)}=(S\Crnd)^t\ket{\psi(0)}.
\end{align}
We chose each random unitary matrix $\Crnd_x$  from the ensemble given in Ref.~\cite{RandomUnitary}.

Let us now define a non-Hermitian extension of the quantum walk.
We make either $S$ or $C$ non-unitary.
The lattice Hamiltonian~\eqref{eq20} of the Hatano-Nelson model inspires us to modify the shift operator in the following way:
\begin{align}
S(g)\ket{xL}&=e^{-g}\ket{(x-1)L},
\\
S(g)\ket{xR}&=e^g\ket{(x+1)R}
\end{align}
for any $x$.
This is actually related to the $PT$-symmetric quantum walk defined in Ref.~\cite{Mochizuki16,Mochizuki20b,Kawasaki20} with a gain-loss operator.
The simplest version of the time-evolution operator defined there is given by
\begin{align}\label{eq170}
U_{PT}(g):=S(0)G(-g)CS(0)G(g)C,
\end{align}
where the gain-loss operator is defined by
\begin{align}
G(g)\ket{xL}&=e^{-g}\ket{xL},
\\
G(g)\ket{xR}&=e^{g}\ket{xR}
\end{align}
for any $x$ while $C$ is set to Eq.~(\ref{eq:Csimple}), for example.
It is straightforward to find that $S(0)G(g)=S(g)$.
We therefore conclude that the time-evolution operator~\eqref{eq170} of the $PT$-symmetric quantum walk is written as
\begin{align}
U_{PT}(g)=S(-g)CS(g)C.
\end{align}
The $PT$-symmetric quantum walk has been experimentally realized by using  classical laser lights~\cite{Regensburger12} and  optical devices with single photons~\cite{Xiao17}.

In the following, we rather focus on the single non-unitary shift operator $S(g)$ with a site-random unitary coin operator $\Crnd$ in Eq.~\eqref{eq290},
\begin{align}
U_\textrm{HN}(g):= S(g) \Crnd.
\end{align}
Below we will show that $U_\textrm{HN}(g)$ has properties similar to the non-Hermitian Hatano-Nelson model reviewed in Sec.~\ref{sec2}.
The most significant difference lies in the fact that the quantum walk exhibits the localization-delocalization transition for any energy eigenstates at the same value of the non-Hermitian parameter $g$.
This is because the quantum walk does not possess any symmetry that is relevant to the classification of topological phases.

\begin{figure}
\centering
\includegraphics[width=0.75\textwidth]{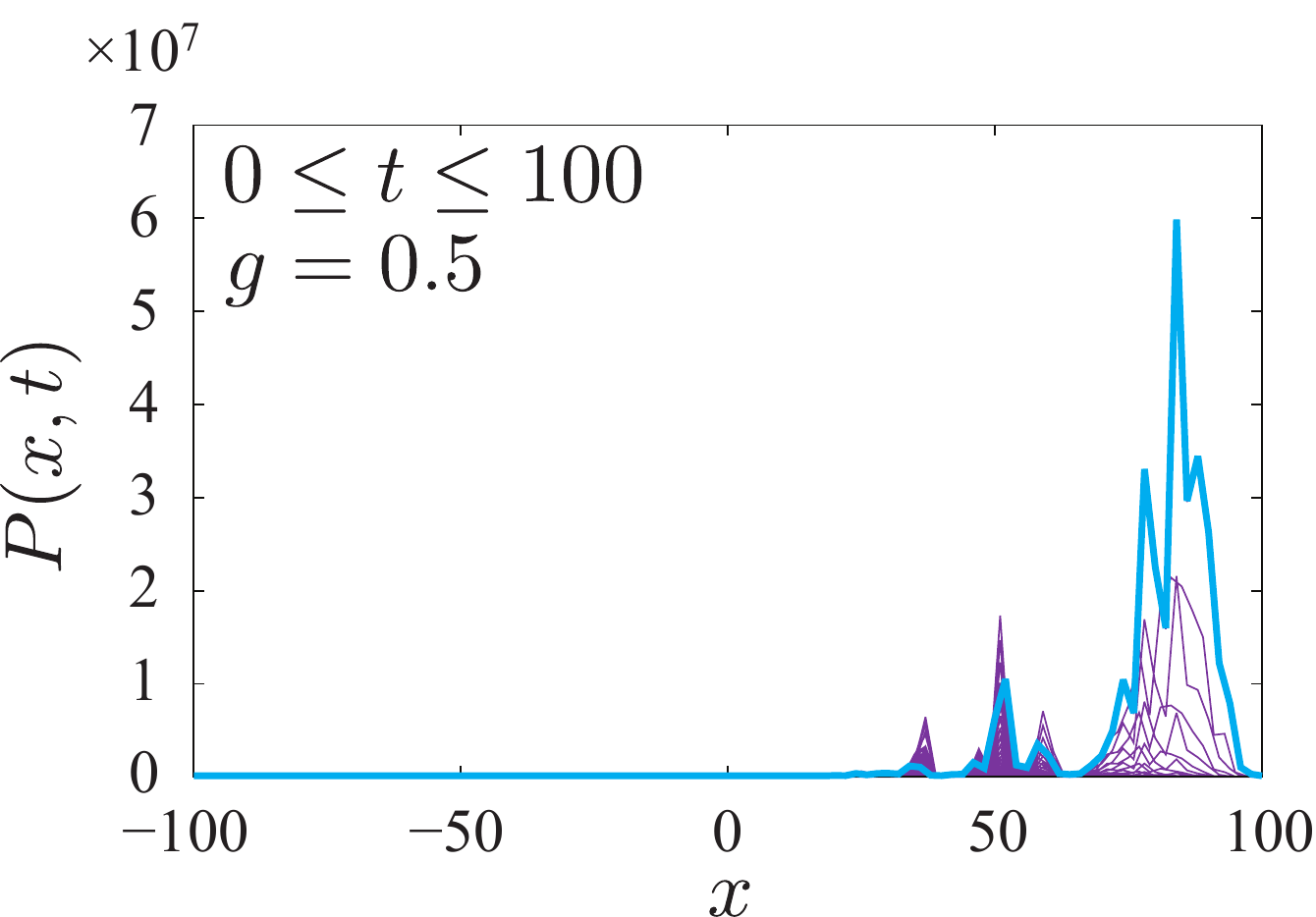}
\caption{The same as in Fig.~\ref{fig4}(b) but with $g=0.5$.}
\label{fig5}
\end{figure}

Figure~\ref{fig5} exhibits what happens to the situation of Fig.~\ref{fig4}(b) when we turn on the non-Hermitian parameter $g$.
We can observe that the probability amplitude moves to the right as the time progresses.
The movement reminds us of variable-range hopping of electrons in disordered semiconductors~\cite{Shklovskii84,Nelson93,HN97};
the quantum walker does not move uniformly, but rather hops from a meta-stable point to the next one.

\begin{figure}
\centering
\includegraphics[width=0.7\textwidth]{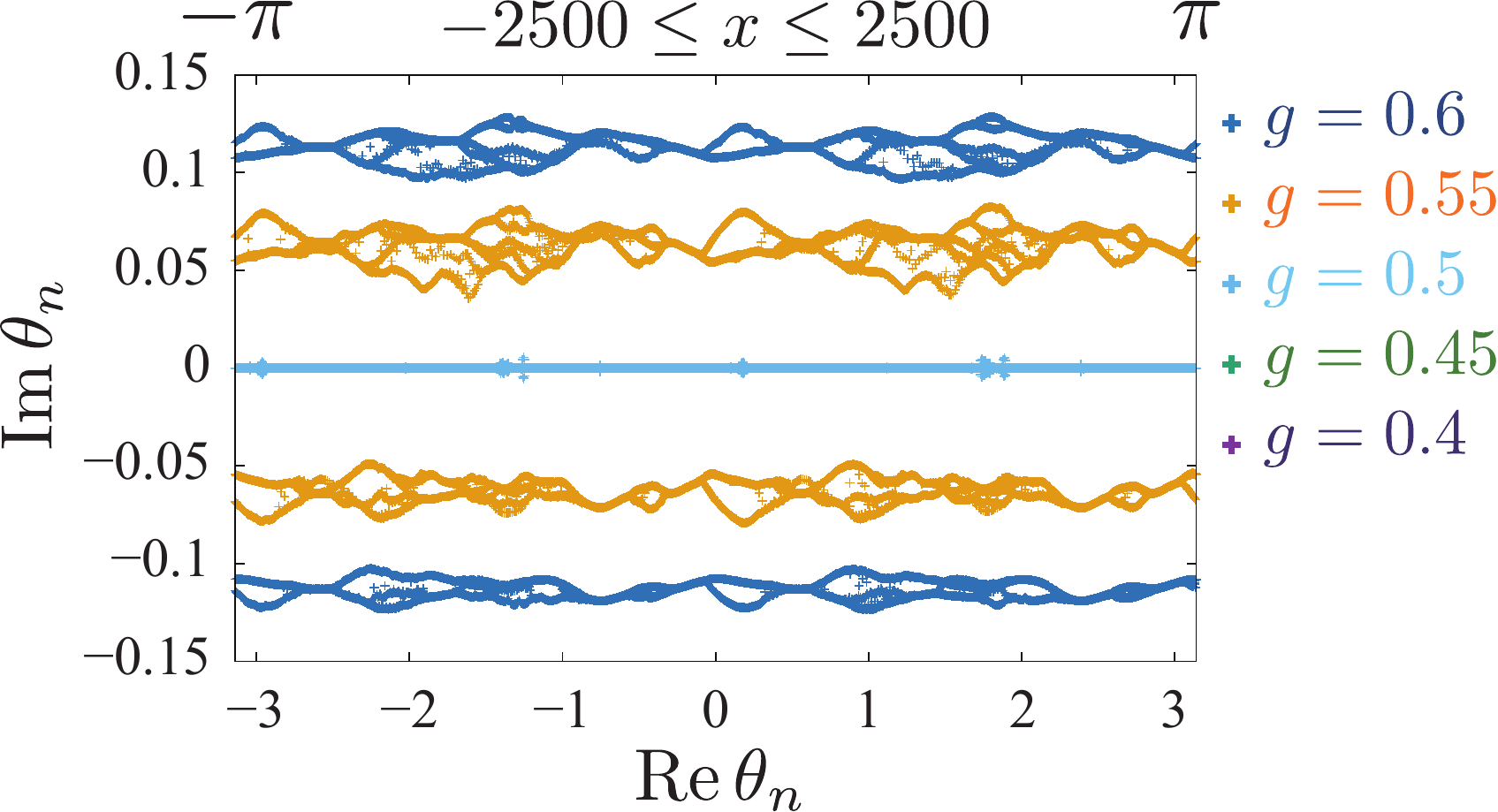}
\caption{The energy eigenvalues of $H(g)$ for $g=0.4$, $0.45$, $0.5$, $0.55$, and $0.6$.
The data points for the eigenvalues for $g=0.4$ and $0.45$ are hidden below those for $g=0.5$.
The length of the system is $5001$.}
\label{fig6}
\end{figure}

Remembering the argument for the Hatano-Nelson model that we summarized near the end of Sec.~\ref{sec2}, we presume that the delocalization transition happens at the same time as the transition in which the energy eigenvalues become complex.
We here define the energy eigenvalues by the eigenvalues of $H(g)=i\log \qty(S(g)\Crnd)$;
in other words, we first numerically find the eigenvalues $\{\lambda_n\}$ of the operator $S(g)\Crnd$  and obtain its phases $\theta_n$ as in $\lambda_n=\exp(i\theta_n)$, where we adjusted the range of its real part as in $-\pi\le\Re\theta_n\le\pi$.
We plotted in Fig.~\ref{fig6} the eigenvalues, varying the non-Hermitian parameter $g$.
We can see here that (i) all eigenvalues behave similarly and (ii) the eigenvalues are on the verge of becoming complex at $g=0.5$.
We thereby presume that all eigenstates of the Hermitian quantum walk with the present unitary randomness have a common inverse localization length, which is close to $\kappa_n\simeq 0.5$. 

Let us confirm this by an independent method of calculation.
We calculate the inverse localization length $\kappa$ of the Hermitian quantum walk by the transfer-matrix method. 
By rearraning the wave function amplitudes around the position $x$ of the eigenvalue equation $U_\textrm{HN}(0) \ket{\psi} = e^{i\theta}\ket{\psi}$, we derive the transfer matrix
\begin{align}\label{eq390}
 \begin{pmatrix}
  \psi_{{x+1},R} \\\psi_{{x},L}
 \end{pmatrix}
&= T_x
 \begin{pmatrix}
  \psi_{{x},R} \\\psi_{{x-1},L}
 \end{pmatrix},
\end{align}
where $\psi_{x,R} = \braket{x,R}{\psi}$ and $\psi_{x,L} = \braket{ x,L}{\psi}$
while the transfer matrix is given by
\begin{align}
T_x &= 
\begin{pmatrix}
\displaystyle
 \frac{e^{i(-\theta+\phi+\alpha)}}{\cos \vartheta} && -e^{i(\alpha+\beta)}\tan \vartheta \\[10pt]
 -e^{i(\alpha-\beta)}\tan \vartheta &&  \displaystyle\frac{e^{i(\theta-\phi+\alpha)}}{\cos \vartheta}
\end{pmatrix}.
\end{align}
Applying Eq.~\eqref{eq390} repeatedly, we have the product the random transfer matrix as in
\begin{align}
 \begin{pmatrix}
  \psi_{{N+1},R} \\\psi_{{N},L}
 \end{pmatrix}
&= \Pi_{x=1}^N T_x
 \begin{pmatrix}
  \psi_{{1},R} \\\psi_{{0},L}
 \end{pmatrix}.
\end{align}
We find the inverse localization length $\kappa$ from the Lyapunov exponent of the product of the transfer matrices.

\begin{figure}
\centering
\includegraphics[width=0.7\textwidth]{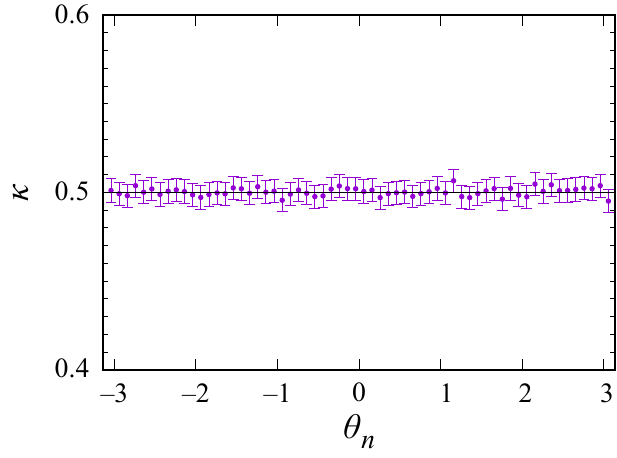}
\caption{The inverse localization length $\kappa$ as a function of the energy $\theta$. The system size is $N=10^5$.}
\label{fig7}
\end{figure}

Figure~\ref{fig7} shows the energy dependence of the inverse localization length $\kappa$ for $N=10^5$. 
In contrast to the result shown in Fig.~\ref{fig3}, the inverse localization length $\kappa$ of $U_\textrm{HN}(0)$ is almost always close to $\kappa=0.5$, not depending on the energy variable $\theta$. This is the reason why all eigenstates of the non-Hermitian quantum walk described by $U_\textrm{HN}(g)$ simultaneously undergo the non-Hermitian delocalization transition when $g=0.5$.

Finally, we argue that the common localization length originates from the absence of symmetry of the quantum walk and the periodicity of the energy eigenvalues without band gaps.
In the Hermitian systems, it is well known that the localization length becomes smaller near band edges; it also diverges at zero energy if the Hamiltonian has chiral and/or particle-hole symmetry. Meanwhile the present quantum walk does not have chiral and/or particle-hole symmetry, because of which the localization length does not diverge at any specific energy eigenvalue.
Furthermore, since the real part of energy eigenvalues of $U_\textrm{HN}(g)$ occupies the whole range of $\Re\theta_n$ in a periodic way as shown in Fig.~\ref{fig6}, there are no band edges. 
These two facts produces the result that the localization length does not depend on the energy eigenvalue.
We remark that since the occupation of the whole range of $\Re\theta_n$ originates from the periodicity of the energy eigenvalue of the quantum walk, the common localization length does not occur in the original Hatano-Nelson model.

\section{Summary}
\label{sec4}

In the present paper, we first review the delocalization transition of the Hatano-Nelson model in one dimension.
At the transition point $g=\kappa_n$, the $n$th eigenvector gets delocalized from a fixed profile of $\braket{\psi^L_n}{x}\braket{x}{\psi^R_n}$ to a nearly plain wave.
At the same time, its eigenvalue jumps out of a fixed real value onto a complex plain.

We then demonstrate that the same delocalization transition occurs for a non-Hermitian extension of the discrete-time quantum walk on a one-dimensional random media.
The result suggests that if we introduce the site randomness by choosing the coin operator for each site out of a random unitary ensemble in Ref.~\cite{RandomUnitary}, all eigenstates have a common value of the inverse localization length.
We indeed confirm this by the transfer-matrix calculation of the Hermitian quantum walk.

\section*{Acknowledgements}

The present authors' work is supported by JSPS KAKENHI Grant Numbers JP19H00658 and JP21H01005.
N.H.'s work is also supported by JSPS KAKENHI Grant Number JP19F19321.
H.O.'s work is also supported by JSPS KAKENHI Grant Numbers JP18H01140 and JP20H01828.


\bibliography{hatano}

\end{document}